\documentclass[prd,preprint,amsmath,nofootinbib,superscriptaddress]{revtex4-1}
\usepackage{setspace}
\usepackage{graphicx,youngtab}
\usepackage{bm}
\usepackage{epsfig}
\usepackage{color}
\usepackage{colordvi}
\newcommand{\nn}{\nonumber}

\newcommand{\beq}{\begin{equation}}
\newcommand{\eeq}{\end{equation}}
\newcommand{\bea}{\begin{eqnarray}}
\newcommand{\eea}{\end{eqnarray}}

\begin{document}

\title{ Partition Functions on the Euclidean Plane with Compact Boundaries in Conformal and Non-Conformal Theories }

\author{ Ira Z. Rothstein } 
\affiliation{%
  Department of Physics, Carnegie Mellon University, 5000 Forbes Ave.,
  Pittsburgh, PA 15213, USA }%

\date{\today}

\begin{abstract}
In this letter we calculate the exact partition function for free bosons on the plane with lacunae. First the partition function for a plane with two spherical holes is calculated by matching exactly for the infinite set of Wilson coefficients in an effective world line theory and then performing the ensuing Gaussian integration. The partition is then re-calculated using conformal field theory techniques, and the equality of the two results is made manifest. It is then demonstrated that there is an exact correspondence between the Wilson coefficients (susceptabilities) in the effective field theory and the weights of the individual excitations of the closed string coherent state on the boundary. We calculate the partition function for the case of three holes where CFT techniques necessitate a closed form for the map from the corresponding closed string pants diagrams. Finally, it is shown that the Wilson coefficients for the case of quartic and higher order kernels, where standard CFT techniques are no longer applicable, can also be completely determined. These techniques can also be applied to the case of non-trivial central charges.\end{abstract}

\maketitle 

\section{ Introduction}

The partition functions for conformal field theories  on surfaces with boundaries are
of interest in diverse areas of physics. 
   In soft condensed matter physics it is of interest to calculate partition
functions on the Euclidean  plane  with compact boundaries for the purpose of understanding fluctuation induced forces \cite{kardar}.
 In string theories, amplitudes correspond  to such partition functions  and one must sum over all moduli
 and Euler characteristics. However, to calculate  a particular gauge slice
is not always simple. Powerful methodologies in boundary conformal field theory (BCFT)
 \cite{Cardy1} have been utilized to determine the spectrum of primary operators in Gaussian as well as non-Gaussian   theories.
 These methods utilize the state operator correspondence to implement the boundary conditions
 via the insertion of  boundary operators.

In this paper  we will take a different approach based on world-line effective field theories (EFT)\cite{GnR} as applied
to statistical systems \cite{YRD1,eisenriegler}. 
In particular,  we will be interested in the case of partition functions on the Euclidean plane with
compact boundaries. The EFT approach is similar to  BCFT  in that 
the effects of the boundaries are accounted for via operator insertions.  However, we will make no use of
the state operators correspondence, and as such, the technique will not be limited to conformal field theories.

We will begin by considering 
the free bosonic action
\begin{equation}
S_0^B =\frac{1}{2} \, \! \int d^2r\,\partial\phi
 \partial \phi
  \ , \label{H}
\end{equation}
on a plane with holes which satisfy
Neumann boundary conditions. This problem is identical to finding the partition function for
a film embedded with rigid disks which are allowed to bob but  not tilt\footnote{This is not true for all observables but will hold for the partition source-free partition function.}, 
where the notion of tilting is given meaning by a particular choice of
gauge. In particular the scalar field partition function will correspond to
the Monge gauge choice \cite{CMTtext} where the height of the field $\phi$ is measured with respect to a Euclidean flat base plane.  In the film model there are non-linear corrections
which are there as a consequence of reparameterization invariance, but it can 
be shown that these corrections are suppressed \cite{YRD1} by powers of $k/ \beta$
where $k$ is the modulus characterizing the surface energy.
The fluctuation induced forces  in these systems are  relevant to soft condensed
matter problems \cite{kardar}.

In \cite{YRD1} it was shown that it is possible to utilize the world-line effective field theory
developed for studying black hole inspirals \cite{GnR} to study the Casimir forces on two dimensional surfaces (the dimensionality is actually irrelevant).  The idea is to work in the point particle
approximation, and then account for the finite size (read boundary conditions) by
adding an infinite tower of higher dimensional operators to the action.
For defects with only linear responses to external perturbations   these terms are
all quadratic in the field, and their dimensionality is fixed by the number of
derivatives. Power counting shows that non-linear interactions on the world-line
generate corrections which are suppressed at high temperatures.
 In  the limit where the size of the
objects ($R$) is small compared to the distance between them $r$, truncating the
infinite sum induces errors of order $(R/r)^\tau$, where $\tau$ is the the number of
derivatives in the first term dropped in the series. 
The coefficients of these operators (Wilson coefficients)  correspond
to generalized polarizabilities and are fixed by performing a matching calculation
between the full theory and the effective theory.  The full set of Wilson coefficients can be calculated
 by first matching the response of  the individual
 holes to multipole fields in the full and effective theory. We will then show that these matching
 coefficients can equally well be extracted by determining the boundary
 operators in the Hamiltonian version of the theory.
Comparing the two procedures
 illuminates a map between closed string excitations and effective polarizabilities.
The model is then solved by summing the series for the higher dimensional
operators into a non-local operator which can be used to 
write down a closed form expression for the partition function.

In the effective theory the action for  for n-bodies may be written as 
\beq
S=S_0+\frac{1}{2}\sum_{i} C_i (\partial_{a_1}....\partial_{a_n} \phi)(\partial_{a_1}....\partial_{a_n} \phi),
\eeq
where the second term is localized at the origin.
Operators involving traces over indices are redundant in this theory as they
can be traded for other operators via a canonical transformation. 
For non-spherically symmetric objects the polarizibilities will transform
non-trivially under the two dimensional rotation group, but are singlets
for the spherical  case of interest here. These coefficients are fixed by a simple
matching procedure whereby we place the objects in a background
field and calculate its response.  For a given polarizability of a given
order 
we may extract the corresponding $C_i$ by placing the object in a multipole
field of the same order. 

To utilize holomorphy we match in the complex plane a general n-pole  background
field takes the form
\beq
\label{bg}
\phi_{bg}(z,\bar z)= \frac{\lambda}{R^n} (z^n+\bar z^n).
\eeq
We will impose Neumann boundary conditions which 
preserve the conformal symmetry.
\beq
\partial_r \phi(z)\mid_{R}=0
\eeq
  This choice means that
the first non-zero $C_i$ will be $C_2$, a dipole polarizability.  
In the full theory we solve the rudimentary boundary value problem.
The induced field is given by
\beq
\delta\phi= \lambda  R^{n}(\frac{1}{z^n} +\frac{1}{\bar z^n}).
\eeq

Expanding around the background field (\ref{bg}) 
utilizing the Greens function in complex coordinates
\beq
G(z,\bar z) = \frac{1}{4 \pi } (\log (z) +\log (\bar z))
\eeq
and calculating the one point
function leads to the EFT result
\beq
\delta \phi =-\frac{C_n \lambda}{4 \pi  R^{n} } 2^{2n-1} n! (n-1)!(\frac{1}{z^{n}}+\frac{1}{\bar z ^{n}}).
\eeq
We may then extract the matching coefficient to be 
\beq
\label{Cs}
C_n= -\frac{\pi }{n! (n-1)!~2^{n-2}}R^{2n}.
\eeq

This form of matching coefficient can also be obtained  from perspective of closed string
theory. If we conformally map an isolated hole  into an infinite cylinder, and consider
the direction along the length as being time-like, then  the partition function corresponds  to a closed string propagator
emanating from a  boundary state \cite{callan}.
The boundary states are annihilated by the constraint
\beq
\partial_1 \phi \mid \psi \rangle =0
\eeq
where the world-sheet is parameterized by $(\sigma_1,\sigma_2)$ and $\sigma_1$ is the
time like direction. 

Our gauge choice (Monge) corresponds to choosing mode oscillators in the string expansion to be  along one space like direction in the embedding space.
 In terms of the mode operators the constraint \cite{callan}reads
\beq
(\alpha_n+\tilde  \alpha_{-n} )\mid \psi \rangle =0
\eeq
where the (standard) normalization is fixed via
\bea
\partial \phi(z) &=& -i \left( \frac{\alpha^\prime}{2}\right)^{1/2}\sum_{m=-\infty}^{\infty} \frac{\alpha_m}{z^{m+1}},\eea 
\beq
[\alpha_m,\alpha_n]=[\tilde \alpha_m,\tilde \alpha_n]=m \delta_{m,-n},
\eeq
and 
\beq
S=\int \frac{1}{4\pi \alpha^\prime} (\partial \phi \bar \partial \phi) d^2 \sigma.
\eeq

The solution to this constraint is
\beq
\mid \psi \rangle  \propto  e^{-\sum_{p=1}^\infty (\frac{1}{p} \alpha_{-p} \tilde \alpha_{-p})}\mid 0 \rangle.
\eeq
Using the fact that
\bea
\alpha_{-m} &=& \left( \frac{2}{\alpha^\prime}\right)^{1/2} \frac{i}{(m-1)!}\partial^m \phi(0)
\eea
the state is written as
\beq
\mid \psi \rangle  \propto  e^{-\sum_{p=1}^\infty \frac{2}{\alpha^\prime p! (p-1)!}(\partial^p\phi \bar \partial^p \phi)}\mid 0 \rangle.
\eeq
rescaling the fields so that they are normalized according to (\ref{H}) leads to 
\beq
\mid \psi \rangle  \propto Exp\left[{-\sum_{p=1}^\infty \frac{4\pi  R^{2p}}{ p! (p-1)!}(\partial^p\phi \bar \partial^p \phi)}\right]\mid 0 \rangle.
\eeq
Taking this as an operator insertion
in the path integral 
reproduces the Wilson coefficients $C_n$. It is interesting to note that there is a
correspondence between the string excitation and the generalized susceptibilities.
For instance the dipole polarizability $(\partial \phi)^2$ on the world line maps to the dilaton excitation of the string. Higher order multipoles map to massive string excitations.

Given the action the two body problem can be solved.
We utilize the holomorphic nature of the fields to re-write the action  as
\bea
S_{int}
&=& \sum_n C_n (2)^{n}(\partial_{z}^n \phi ) (\partial_{\bar z}^n \phi )\nn \\
\eea
Then using the result  (\ref{Cs}) we have
\bea
&&S=\int d^2z  [\frac{1}{2}\partial_z \phi \partial_{\bar z} \phi+ 
\sum_a  (4 \pi )  R_a \sqrt { \partial_z \partial_{\bar z}}I_1(2R_a \sqrt { \partial_z \partial_{\bar z}})\phi(z) \phi(\bar z)\delta^2(z-z_a,\bar z - \bar z_a)].\nn \\
\eea
The exact result for the partition function follows from treating these operators as mass insertions and
recognizing that the contractions between two identical world lines lead to trivial
renormalizations that are pure counter-terms. Thus the only relevant vacuum diagrams are those
in which no two mass insertions are adjacent.

Let us consider the case of two particles.   To calculate the free energy we draw all possible vacuum diagrams
which have no disconnected sub-components.  We must always add two insertions at a time since
for  an odd number of insertions there would have to be a self-energy graph inserted which we renormalize to zero.
Diagrams which are related by a
reflection symmetry are not distinct. 
\begin{figure}[thb]
\centering
\includegraphics[scale=.89]{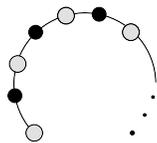}
\caption{All connected vacuum diagrams can be written in term of this
topology. The different colored dots represent insertions at each of the two particles respectively.}
\label{fig1}
\end{figure}
At second order in the interaction the diagram reduces down to a simple bubble 
and the resulting free energy is given by
\bea
\label{exact}
-\beta F  &=&
\sum_{n=1}^{\infty}\prod_{a=1,3,5..}^{2n-1} (R_{A}R_{B})^n\frac{1}{n} \sqrt{ \partial_{z_a}\partial_{\bar z _a}}\sqrt{ \partial_{z_{a+1}}\partial_{\bar z _{a+1}}}I_1(2R_A\sqrt{ \partial_{z_a}\partial_{\bar z_a}})I_1(2R_B\sqrt{ \partial_{z_{a+1}}\partial_{\bar z_{a+1}}}) \nn \\
&\times& \log( z_a- z_{a+1})\log(\bar z_{a}-\bar z_{a+1})\nn \\
\eea
Each $n$ represents two insertions, one of each type, and $z_{2n+1}=z_1$.
Upon expanding this expression and differentiating one sets $z_a=z_1$ and $z_{a+1}=z_2$.
Let us now expand this result in $R/r$ to compare with the literature on
the lower order contributions.
The  leading order  contribution is the dipole-dipole interaction.
Expanding out (\ref{exact}) gives
\bea
-\beta  F^{(2)}_{d-d} =
&=& R_A^2R_B^2  \partial_{z_1}\partial_{\bar z _1} \partial_{z_{2}}\partial_{\bar z _{2}} \log((z_1-z_{2})\log(\bar z_1-\bar z_{2}))\nn \\
&=&\frac{1}{r^4}R_A^2R_B^2 
\eea

We also generate  non-linear terms such as the case  with two dipoles of each type in which case $i=2$.
\bea
-\beta F^{(2)}_{d^2-d^2} = && R_{A}^2R_{B}^2\frac{1}{2} \sqrt{ \partial_{z_1}\partial_{\bar z _1}}\sqrt{ \partial_{z_{2}}\partial_{\bar z _{2}}}I_1(2R_A\sqrt{ \partial_{z_1}\partial_{\bar z_1}})I_1(2R_B\sqrt{ \partial_{z_{2}}\partial_{\bar z_{2}}}) \log(\bar z_1-\bar z_{2})\log(z_1-z_{2})\nn \\
&\times& \sqrt{ \partial_{z_3}\partial_{\bar z _3}}\sqrt{ \partial_{z_{4}}\partial_{\bar z _{4}}}I_1(2R_A\sqrt{ \partial_{z_3}\partial_{\bar z_3}})I_1(2R_B\sqrt{ \partial_{z_{4}}\partial_{\bar z_{4}}}) \log(\bar z_3-\bar z_{4})\log(z_3-z_{4})\nn \\
&=&
R_{A}^4R_{B}^4\frac{1}{2r^8} \nn \\
\eea
All of these perturbative results agree with those in the  literature \cite{GGK,LO}.
New results can be read off by going to higher order.

The case of N holes follows in a similar fashion. If we are interested in the N-body potential   we have a minimum of N insertions and we may add one
vertex insertion at a time. 
\bea
\label{sum}
-\beta F_N=&&\sum_{n=N}^{\infty}\sum_{\cal P}\prod_{a=1}^{n} R_1^{i_1} R_2^{i_2}.... R_N^{i_N} \sqrt{ \partial_{z_a}\partial_{\bar z _{a+1}}}I_1(2R_a\sqrt{ \partial_{z_a}\partial_{\bar z_{a+1}}})
 \times G( z_a- z_{a+1},\bar z_a- \bar z_{a+1})\nn \\
\eea
The second sum is over all possible choice non-equivalent permutation of  vertices.
Where two permutations are in the same equivalence class  if they can be rotated into each other.
 Defining $G(0)=0$ implies that all choices with two adjacent
identical vertices vanish. For each choice of vertices,  $(i_1,i_2,i_3...)$ correspond to the number of vertices of (hole) type ($1,2,3...$) in the
particular term in the sum. The expression (\ref{sum}) is easily evaluated to arbitrary order in MATHEMATICA, though
this particular form is most probably not the most efficient since it entails redundancies as compared to the two body 
results presented above.


\section{CFT Solution for Two Holes}
The determinant of the Laplacian on Riemann surfaces  in many instances can recovered
by conformally mapping surfaces for which the partition is easily calculated to surfaces
which present more of a challenge, such as the case of interest in this paper.
The techniques used here are well known though it seems that the  explicit result for the case of interest, two spheres in a plane, is absent in the voluminous literature
on the subject of functional determinants in two dimensions.

The partition function on the plane with two holes can be calculated by a series of conformal
maps starting from the cylinder. One could of course map directly from the cylinder
to the plane with holes, but it is perhaps simpler to use the annulus as an intermediate step.
In each step we work in a conformal gauge 
\beq
g_{ab} \sim e^{2 \phi} \hat g_{ab},
\eeq
in which case the following quantity is an invariant  \cite{alvarez,weisberger}
\beq
I_N
= \frac{Det^\prime_N(\nabla^2)}{A}( e^{\frac{L}{6\pi} -\frac{1}{4\pi} \int_{\partial M} \kappa ds}).\eeq
 Where we have chosen Neumann boundary conditions for the purpose of illustration.
 $L$ is defined as 
\beq
L(\phi,\hat g)= \int_M d\hat \mu \hat K \phi +\int_{\partial M} d\hat s \hat \kappa \phi+\frac{1}{2} \int_M d\hat \mu \hat g^{ab}(\partial_ a \phi \partial_b \phi),
\eeq 
and  hatted quantities are defined by the hatted metric, $K$ is the Gaussian curvature on $M$, $\kappa$ is the geodesic curvature on the boundary, and $A$ is the area of $M$. The prime above the determinant denotes the fact that 
the zero mode is excluded.
 
Beginning with the partition function on a cylinder with length $l$ and unit radius. A rudimentary calculation yields the  result\beq
Z= \frac{1}{\sqrt{2l}} e^{\frac{\pi l}{6}} \prod_{n=1}^\infty \frac{1}{1- e^{-4\pi l n}}.
\eeq
We then map this result to the annulus using the conformal transformation
\beq
z= r_2 e^{2 \pi i w}
\eeq
with 
\beq
2\pi l = \ln(r_2/r_1).
\eeq
The outer (inner) radii of the annulus are $r_2 (r_1)$. The conformal factor is given by
\beq
\phi= \frac{1}{2}\log( 2 \pi z \bar z).
\eeq
Given that both spaces are flat we find
\beq
L_{ann}= \pi \log (r_2/r_1).
\eeq
The annulus partition function is then given by \cite{weisberger}
\beq
Z_{ann} = \frac{f(r_1^2/r_2^2)}{\sqrt{2(r_2^2-r_1^2)}} (r_1/r_2)^{-1/6}
\eeq
\beq
f(x)= \prod_{n=1}^\infty\frac{1}{1-x^{n}}
\eeq
Then we can arrive at case of interest  (z plane with two holes)  via a conformal map from the annulus (w plane).
\begin{figure}[thb]
\centering
\includegraphics[scale=.93]{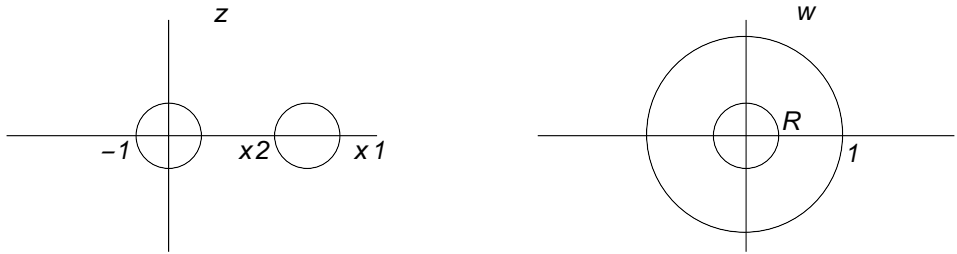}
\caption{Mapping between annulus and twice punctured plane}
\label{map}
\end{figure}
Tha mapping   shown in  figure (\ref{map})  is given by
\beq
\label{Cmap}
w=\frac{z-a}{a z-1}
\eeq

 where
 \beq
 a= \frac{1+x_1 x_2 +\sqrt{(x_1^2-1)(x_2^2-1)}}{x_1+x_2}
 \eeq
 \beq
 R= \frac{x_1 x_2 -1 -\sqrt{(x_1^2-1)(x_2^2-1)}}{x_1-x_2},
 \eeq
under the conditions
$x_2<a<x_1$.

 The metric on the  annulus is then mapped to the flat z plane
 \beq
  dz d\bar z=d w d\bar w \frac{(a^2-1)^2}{(aw-1)^2 (a \bar w-1)^2}
 \eeq
 so that
 \beq
 \phi=\frac{1}{2} \ln \left(  \frac{(a^2-1)^2}{(aw-1)^2 (a \bar w-1)^2} \right)
 \eeq
 and $\hat g=\delta_{ab}$ is the metric on the flat plane so that $\hat K=0$. 
 The annulus is flat ($R \sim \nabla^2 \phi$) so $\int_{\partial M} \kappa =0$ since its Euler characteristic  $\chi=0$.
 However, the integral
 \beq
 \int_{\partial M} d\hat s \hat \kappa \phi=\int_0^{2\pi} d \theta (\phi(1,\theta)-\phi(R,\theta))= 2 \pi Log[R] 
 \eeq
 does not vanish since $\phi$ is not constant along the boundary.
 
 The bulk contribution to the anomaly must vanish since the conformal map is an element of the $SL(2,C)$ subgroup
 with vanishing Schwarztian. Hence we are left the the following partition function for the plane with two holes.
 \beq
 \label{CFT}
 -\beta F= Log[ f[ R^2]]
 \eeq
 where
 \beq
f[x]\equiv  \prod_{n=1}^{\infty}\frac{1}{ (1- x^n)}
 \eeq
and
 \beq
 R= 1/2(r^2-r \sqrt{r^2-4} -2).
 \eeq 
 Expanding in the ratio of the size of the holes relative to the distance between them gives
 \beq
 \beta F \approx \frac{1}{r^4} + \frac{4}{r^6}+\frac{31}{2r^8 }+\frac{60}{r^{10}}+...
 \eeq
 Which agrees with the expansion of (\ref{exact}).
 The expression (\ref{CFT}) certainly has many advantages over the result (\ref{exact}). In particular, though the
 expansion in (\ref{exact}) has a finite radius of convergence we can not expand it around to
 the osculating limit. Whereas  the Dedekind function which results from (\ref{CFT},41) has multiple
 representations which can be easily expanded around this limit. The result (\ref{CFT}) may be used
 be compared to
 the so-called ``proximity force approximation''\cite{NPA} which has recently been systematized in
 terms of a derivative expansion \cite{systematic}. 
 \section{Non-Conformal Theories}
It is interesting to see that  the effective theory methodology can still
 yield exact results even for non-conformal kernels.  A non-conformally invariant  kernel of particular interest is the bi-Laplacian which
 corresponds to a model of a membrane where bending dominates tension.
  In this case there are more operators at each order in $R$
 since the equations of motion can no longer be used to eliminate operators involving a single Laplacian.
In the case of the
bi-Laplacian, beyond second order in derivatives 
 we have two operators at each order in the derivative expansion. 
At  the quadrapole  level we have two possible
operators. 
 A convenient basis to choose is
\beq
O^{2}_a= \partial_z \partial_z \phi \partial_{\bar z} \partial_{\bar z} \phi,~~~O^{2}_b=  \partial_z \partial_{\bar z} \phi \partial_{\bar z} \partial_{ z} \phi.
\eeq


At higher orders we add one power of $\partial_{z,\bar z}$ to each side. So that at $n'th$  order in derivatives we have
\beq
O^{n+1}_a= \partial^{n+1}_z \phi  \partial_{\bar z}^{n+1}\phi,~~~O^{n+1}_b=  \partial_z^n \partial_{\bar z}\phi \partial_{\bar z} ^n\partial_{ z} \phi.
\eeq
In the case of the bi-Laplacian one must impose two boundary conditions to find unique
solutions.  Which operator has the first non-vanishing susceptibility will depend upon the particular choice.
The crucial point is that the number of operators at each level in derivatives does not grow which allows
again for  the matching procedure to carried out to all order in the derivative expansion \cite{YD}.

This fact may at first seem surprising given that the theory is not conformal. However,  the theory is invariant under the 
$SL(2,C)$  sub-group of the Virasoro group  which is revealed once one makes a field redefinition \cite{Bludman,Ferrari}
\beq
\phi \rightarrow \mid\! \frac{dz}{dz^\prime} \!\mid \phi^\prime (z^\prime).
\eeq 
In addition, higher order kernels can also be solved by the same method. However, as one
goes to higher orders more (but still finite) operators appear at each order.


\section{Conclusions}
Partition functions for on  higher genus surfaces  can have a rich structure.
The boundaries introduce new sets of operators which make even free theories
highly non-trivial. 
 Models which are integrable on simple manifolds, such as
cylinders,  can be solved on more complicated surfaces through conformal mappings and
the associated anomalous transformations of the stress energy in the bulk and boundary \footnote{The partition function for sections of  spheres have been calculated in this fashion in \cite{dowker}}.
However, mappings to the Euclidean plane with multiple boundaries of various shapes
can become analytically intractable. 

In this letter we have shown that two dimensional  partition functions 
can also be expressed in terms of an infinite set of Wilson coefficients in a world line EFT. For Gaussian theories, such as the free bosonic one discussed here, these coefficients are exactly calculable. These coefficients which physically correspond
to generalized susceptibilities, are equal to the weights of string excitations in the coherent state 
boundary operator in the closed string channel. 
Given these coefficients, the partition function follows from the vacuum energy graphs. The result is 
written as an infinite sum which is clearly inferior to  the infinite product
form which arise in the context of CFT methodology. 
For the case of two spherical holes we have shown the equality of the CFT and EFT results expanding to arbitrary order in the separation
between the holes.

  However, as the number of 
defects increases the only impediment to solving the theory using the world-line methodology is that the  series involved
grows more cumbersome. Here we only presented the sums in their most raw form,
and it is surely possible to simplify them further \cite{YRD}. Furthermore, it is also possible
to determine the Wilson coefficients for deformed holes using the fact that
can take advantage of rudimentary conformal transformation techniques to calculate
the infinite set of Wilson coefficients for ellipses  \cite{Haussman}.
We have also shown that  it is also possible to solve for the Wilson  coefficients for the  non-conformal bi-Laplacian kernel.

Finally it is interesting to ask how these EFT techniques can be applied to interacting theories.
For any conformal theory, the conformal mapping technique is still applicable, and there is
no reason that one could not find the fluctuation induced force say for an Ising model \cite{Eisenriegler}.
However, again if one is interested in more complicated geometries (such as multiple holes) the mappings will
become analytically intractable. However, given that distortions of single holes to other shapes, such as
ellipses, is fairly simple, the matching procedure in the effective theory technique is still relatively straightforward.
Thus, given that for a simple geometry one can solve for the Wilson coefficient, at least for an integrable
model, one can also determine the coefficients for more complicated shapes. The free energy for
multiple holes can then be calculated via the vacuum energy diagrams as shown above, at least for
 theories which are integrable.


\begin{acknowledgments}
This work  is supported by US DOE contract, 22645.1.1110173.
I thank the Caltech high energy group for its hospitality
and  the Gordon and Betty Moore Foundation for support. 
I benefited from conversations with Anton Kapustin and Rich Holman, Markus Deserno and Cem Yolcu. I thank W. Goldberger for 
comments on the manuscript and Erich Eisenriegler for pointing me to \cite{Eisenriegler}.
\end{acknowledgments}


\end{document}